# Microwave response of bulk MgB$_2$ samples of different granularity

A. Agliolo Gallitto [a], G. Bonsignore [a], G. Giunchi [b], M. Li Vigni [a], Yu. A. Nefyodov [c]

[a] Dipartimento di Scienze Fisiche ed Astronomiche, Università di Palermo, via Archirafi 36, I-90123 Palermo, Italy

[b] EDISON S.p.A., Divisione Ricerca e Sviluppo, Foro Buonaparte 31, I-20121 Milano, Italy

[c] Institute of Solid State Physics RAS, 142432 Chernogolovka, Moscow, Russia

agliolo@fisica.unipa.it

**Abstract**. The microwave response of three high-density bulk MgB$_2$ samples has been investigated in the linear and nonlinear regimes. The three samples, characterized by different mean size of grains, have been obtained by reactive infiltration of liquid Mg in powdered B preforms. The linear response has been studied by measuring the microwave surface impedance; the nonlinear response by detecting the power radiated by the sample at the second-harmonic frequency of the driving field. Our results suggest that bulk MgB$_2$ prepared by the liquid Mg infiltration technique is particularly promising for manufacturing resonant cavities operating at microwave frequencies.

## 1. Introduction

Because of the reduced energy losses, superconductors (SC) are particularly recommended to manufacture components operating at microwave (*mw*) frequencies [1]. Particular attention has been devoted to design superconducting resonant cavities with very high quality factor, which can be conveniently used for both basic and applied research, e.g., for particle accelerators. In spite of the high $T_c$ of cuprate SC, superconducting cavities are currently fabricated by using bulk and/or film of Nb, and its alloys; a further promising candidate for this purpose seems to be the magnesium diboride [2]. The advantages of using MgB$_2$ are several: the relatively high critical temperature; the long coherence length, which makes the materials less susceptible to structural defects like grain boundaries; the ductility and malleability of the compound, related to its metallic nature.

At present, one of the main factors limiting the use of SC in *mw* devices is the presence of nonlinear effects which arise at high input power; so, it is of great relevance to establish the mechanisms responsible for the nonlinear response and the specific properties of superconducting samples to be conveniently used in manufacturing *mw* devices. Nonlinear effects bring about input power dependence of the surface impedance, intermodulation product and harmonic emission [3]. Because of the very high power levels employed in particle accelerators, nonlinear effects develop, worsening the performance of the accelerating line.

In this work we have studied the *mw* response of three high-density samples of bulk $MgB_2$, in the linear and nonlinear regimes. The three samples, characterized by different mean size of grains, have been obtained by reactive infiltration of liquid Mg in powdered B preforms. The investigation in the linear regime has been carried out aiming to determine specific properties of the samples, related to the presence of defects and/or inhomogeneity, such as the residual surface resistance at low temperature, $R_{res}$, and the residual resistivity ratio, $RRR \equiv \rho(300 \text{ K})/\rho(T_c)$. In the nonlinear regime, we have investigated the second-harmonic emission at low static magnetic fields, when the nonlinear processes arising from motion of Abrikosov fluxons are not effective. In this way, we explore the effects of weak links in the nonlinear response of the different samples.

## 2. Samples

The $MgB_2$ samples were prepared by the reactive liquid Mg infiltration technology [4], which consists in the reaction of boron powder (commercial purity, Stark H. C., Germany) and pure liquid Mg, inserted in a Stainless Steel container, with a weight ratio Mg/B over the stoichiometric value ($\approx 0.55$). The container was sealed by conventional Tungsten Inert Gas welding procedure, with some air trapped inside the boron powder.

To obtain the materials from which samples #1, #2 and #3 have been extracted we performed three separate preparations, using different boron powders: a) for sample #1, we used micron size amorphous B powder (98% purity, Grade I) and a thermal annealing at 900°C for 30 min; b) for sample #2, we used crystalline B powder (99.5% purity, original chunks mechanically grinded and sieved under a 40 micron sieve) and a thermal annealing at 900°C for 1 h; c) for sample #3, we used crystalline B powder (99.5% purity, original chunks mechanically grinded and sieved under a 100 micron sieve) and a thermal annealing at 900°C for 2 h. All the $MgB_2$ samples, after reaction, reach very high densities (2.40, 2.35 and 2.30 g/cm$^3$ for samples #1, #2 and #3, respectively). As previously reported for analogous $MgB_2$ products, the size of the used boron powder affects greatly the morphology and the superconducting characteristics of the final material; in particular, sample #1 shows better transport and magnetic properties [5,6]. The samples have approximate dimensions $2 \times 3 \times 0.5$ mm$^3$, their largest faces have been mechanically polished, to obtain very smooth surfaces.

## 3. Experimental Apparatus

The *mw* response has been investigated by two different experimental techniques. In the linear regime (input power of the order of –30 dBm), we have measured the real and imaginary components of the microwave surface impedance, $Z_s = R_s + i X_s$, as a function of the temperature, using the technique of hot-finger cavity perturbation [7]. A cylindrical niobium cavity, resonating in the $TE_{011}$ mode at 9.4 GHz, is maintained in thermal contact with a bath of liquid helium. The sample is inserted into the centre of the cavity by a sapphire rod and insulated from the thermal bath. The temperature of the sample can be varied continuously in the range $4.2 \div 300$ K. Measuring the quality factor and resonance frequency shift of the cavity, we have determined the temperature dependence of $R_s$ and $X_s$.

The measurements in the nonlinear regime (input peak power $\sim 30$ dBm) have been performed by the second-harmonic (SH) generation technique. In this case, the sample is placed in a bimodal cavity, resonating at the two frequencies $\nu$ and $2\nu$, with $\nu \approx 3$ GHz, in a region in which the *mw* magnetic fields $H(\nu)$ and $H(2\nu)$ are maximal and parallel to each other. The $\nu$-mode of the cavity is fed by a train of *mw* pulses (pulse width 5 $\mu$s; pulse repetition rate 200 Hz). The SH signal radiated by the sample is detected by a superheterodyne receiver. The sample is also exposed to a DC magnetic field, $H_0$, parallel to $H(\nu)$ and $H(2\nu)$. In order to highlight possible SH signals related to nonlinear processes in weak links, the attention has been devoted to the SH response at low external fields ($H_0 < H_{c1}$).

## 4. Experimental Results and Discussion

### 4.1. Surface Impedance

As it is well known, the *mw* surface resistance, $R_s$, is related to the energy losses induced by the *mw* current in the surface layer of the sample; it is proportional to $(1/Q_L - 1/Q_U)$, where $Q_U$ is the quality-factor of the empty cavity and $Q_L$ is that of the loaded cavity. The values of $R_s(T)$ in absolute units have been determined taking into account the geometrical factor of the sample, which depends on the sample shape and the configuration of the *mw* fields in the region of the cavity in which the sample is located [7]. The surface reactance, $X_s$, is proportional to the field penetration depth λ; it can be determined measuring the shift of the resonance frequency of the cavity, apart from an additive constant $X_0$. To obtain $X_s$ in absolute units, it is necessary to know $X_0$ that can be determined by imposing the condition of normal skin effect at $T = T_c$, i.e. $X_s(T_c) = R_s(T_c)$. However, our results have highlighted that in the investigated samples the normal skin effect is not verified; so, we report the variation $\Delta X_s(T) \equiv X_s(T) - X_s(4.2\ K)$, which indicates the temperature dependence of λ.

Fig.1 shows $R_s$ and $\Delta X_s$ as a function of the temperature, for the three samples. The inset shows $R_s(T)$ and $\Delta X_s(T)$ in the range $4.2\ K \div T_c/2$: both $R_s$ and $\Delta X_s$ linearly depend on the temperature. The residual surface resistance, $R_{res} \equiv R_s(T \to 0)$, is an important parameter for potential application of superconducting materials; indeed, it marks the energy losses due to impurities and/or inhomogeneity. As one can see from the inset, all the samples exhibit small values of $R_{res}$; in particular, we obtain $R_{res} < 0.5$ mΩ, 4.0 mΩ and 7.5 mΩ, for samples $MgB_2$ #1, #2 and #3, respectively. By using the expression $R_s = (\omega\mu_0\rho/2)^{1/2}$, we have calculated the RRR ratio, obtaining RRR = 4.2, 3.2, 2.7 for $MgB_2$ #1, #2 and #3, respectively. These results show that all the three samples are of high quality and, in particular, the sample with grains of the smallest mean size is that of highest quality.

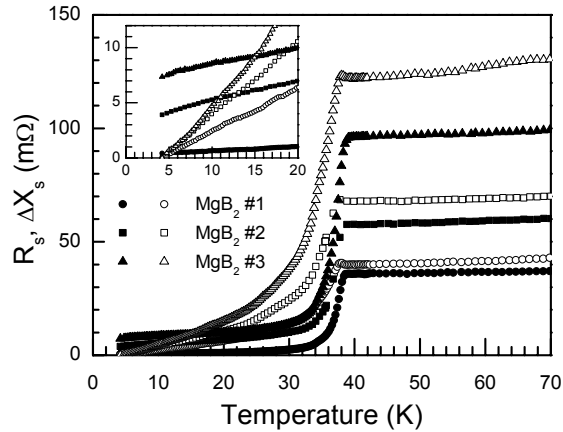

Fig.1. Full symbols: $R_s(T)$; open symbols: $\Delta X_s(T) \equiv X_s(T) - X_s(4.2\ K)$.

### 4.2. Second-Harmonic Emission

Several studies reported in the literature have shown that high-$T_c$ SC are characterized by markedly nonlinear properties, when exposed to intense *em* fields up to *mw* frequencies [3,8,9]. Different mechanisms of harmonic emission have been recognized; in particular, harmonic emission detected at low DC fields and low temperatures has been ascribed to processes occurring in weak links [3,8,9]. Since the first studies performed in $MgB_2$, it has been highlighted that grain boundaries do not appreciably affect the transport properties of this material, in contrast to what occurs in cuprate SC. Nevertheless, measurements of SH signal at low DC fields performed in ceramic samples of $MgB_2$ have shown that at low temperatures nonlinear processes in weak links are the main source of SH emission [9].

Fig.2 shows the SH signal intensity as a function of $H_0$, for the sample $MgB_2$ #3 and a bulk $MgB_2$ sample synthesized by direct reaction of B powder with Mg metal [10], which we indicate as

MgB$_2$ #4. As one can see, the signal exhibits qualitatively the same behaviour for the two samples: it shows a magnetic hysteresis even at $H_0 < H_{c1}$. A peculiarity of the signal is that after $H_0$ reaches values higher than $H_{c1}$, this low field structure disappears irreversibly in successive runs at low fields.

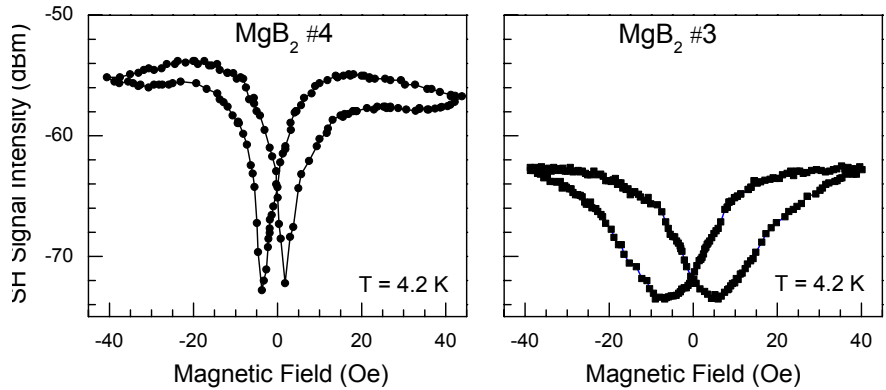

Fig.2. Low-field behaviour of the SH signal. Input peak power ≈33 dBm; noise level −75 dBm.

SH signals exhibiting such butterfly-like hysteresis have been detected in several ceramic samples of MgB$_2$ and also in other kinds of SC [8,9]. The presence of the low-field hysteresis and its irreversible vanishing, after expositing the sample to fields higher than $H_{c1}$, indicate that processes of trapped Josephson fluxons in weak links come in to play in the SH emission [8,9]. As one can see from Fig.2, the signal detected in MgB$_2$ #3 is weaker than that detected in MgB$_2$ #4. Furthermore, we would remark that we have investigated several samples produced by infiltration of liquid Mg in B powder and all the samples have shown very weak SH signals. Concerning the samples investigated in this paper, we have obtained for MgB$_2$ #2 a SH signal of the order of the noise level and not detectable signals in MgB$_2$ #1. These results show that nonlinear processes in weak links are not enough effective to generate noticeable SH emission, especially in the sample with the smallest grain size.

## 5. Conclusion

We have measured the linear and nonlinear *mw* response of three bulk samples of MgB$_2$, produced by reactive liquid Mg infiltration technology, characterized by different mean size of grains. Our results show that the sample with the smallest mean size of grains has better performance. This is probably due to the reduced dimension of the grain boundaries, which makes them less effective in worsening the mw response. The reduced intensity of the SH signals, with respect to those detected in samples prepared by other methods [9], as well as the small value of the residual surface resistance at low temperatures, show that samples produced by the method we used, are particularly promising for manufacturing resonant cavities operating at *mw* frequency.